\documentclass[final,5p,times,twocolumn,number,sort&compress]{elsarticle}

\usepackage[T1]{fontenc}
\usepackage{graphicx}
\usepackage{xcolor}
\usepackage{booktabs}
\usepackage{array}
\usepackage{amsmath}
\usepackage{threeparttable}
\usepackage{siunitx}
\usepackage{tikz}
\usetikzlibrary{arrows.meta,shapes.geometric,calc,backgrounds,fit,matrix}
\usepackage{pgfplots}
\usepackage{verbatim}
\usepgfplotslibrary{statistics}
\pgfplotsset{compat=1.18,compat/show suggested version=false}
\usepackage{hyperref}

\newcolumntype{L}[1]{>{\raggedright\arraybackslash}p{#1}}
\newcommand{\celllines}[1]{\shortstack[l]{#1}}

\hypersetup{
    breaklinks=true,
    colorlinks=false,
    urlcolor=blue,
    hypertexnames=false
}

\journal{Future Generation Computer Systems}
\begin{document}

\begin{frontmatter}

\title{A Cloud Continuum Research Infrastructure for Distributed CPS Experimentation}

\author[bio,messina,cini]{Fabio Orazio Mirto}
\author[messina,cini]{Giuseppe Tricomi}
\author[messina,cini]{Luca D'Agati}
\author[bologna,cini]{Andrea Sabbioni}
\author[icar]{Stefano Silvestri}
\author[messina,cini]{\\Francesco Longo}
\author[messina,cini]{Giovanni Merlino}
\author[bologna,cini]{Armir Bujari}
\author[bologna,cini]{Paolo Bellavista}
\author[messina,cini]{Antonio Puliafito}

\affiliation[bio]{organization={Department of Biomedical, Dental, and Morphological and Functional Imaging Sciences, Università degli Studi di Messina}, country={Italy}}
\affiliation[messina]{organization={Department of Engineering, Università degli Studi di Messina},
            country={Italy}}
\affiliation[bologna]{organization={Università degli Studi di Bologna},
            country={Italy}}
\affiliation[icar]{organization={Institute of ICAR CNR},
            addressline={Naples},
            country={Italy}}
\affiliation[cini]{organization={CINI\@: National Interuniversity Consortium for Informatics}, 
            addressline={Rome}, 
            country={Italy}}

\begingroup
\setlength{\linewidth}{\textwidth}

\endgroup

\begin{abstract}
Cloud Continuum applications require experimental environments capable of combining heterogeneous Edge, Fog, Cloud, and high-performance computing resources while preserving reproducibility, observability, and control over distributed deployments. This paper presents a two-level reference architecture for Cloud Continuum experimentation built on top of the SLICES Cloud Continuum Blueprint. The proposed approach separates the research-infrastructure layer, which exposes and manages distributed resources, from the application layer, where cyber--physical workflows are organized according to an Edge--Fog--Cloud pattern in which placement, timing, and data provenance are treated as first-class experimental concerns.

The architecture is designed to support multiple continuum applications rather than a single domain-specific prototype. At the Edge, applications interact with physical devices and perform low-latency sensing or safety actions; at the Fog, they execute near-source coordination, mediation, and stream-processing logic; at the Cloud, they consolidate global knowledge through analytics, optimization, and visualization.  This partitioning enables researchers to deploy, customize, and compare alternative control and monitoring strategies over the same programmable infrastructure substrate.

The approach is validated through two representative use cases: Renewable Energy Community management, where distributed Digital Twin coordination and time-window-based energy control are requested, and AirWatch, a monitoring pipeline focused on anomaly detection, low-latency alerting, and cloud-side aggregation. Both workloads are evaluated through a systematic campaign of 40 runs comparing virtualized and physical edge deployments over a geographically distributed infrastructure. 
Results show that the same architectural primitives can support both control and monitoring workloads with bounded and attributable performance overhead, providing a workflow-evidence-based basis for experimentally grounded Cloud Continuum research.
\end{abstract}


\end{frontmatter}

\section{Introduction}
\label{sec:intro}

The Cloud Continuum is increasingly recognized as a key execution paradigm for modern distributed applications, especially in Cyber--Physical Systems (CPSs) that combine sensing, local processing, near-edge coordination, and cloud-scale analytics~{\cite{fogrole,edgevision,iotfogcloud}}. In these systems, computation is no longer confined to a centralized data center but is distributed across Edge, Fog, and Cloud resources with different latency, capacity, and reliability characteristics~{\cite{fogiot,cloudtofog}}. This shift creates important opportunities for application design, but it also makes research and experimentation more difficult.

The main challenge is that continuum solutions must be validated over heterogeneous and geographically distributed infrastructures. This experimental space also extends toward High-Performance Computing (HPC) resources, which provide the capacity required for optimization, simulation, model calibration, and data-intensive analytics, while remaining based on execution models and scheduling policies that differ from latency-sensitive Edge--Fog--Cloud deployments. As a result, many proposals are easy to describe conceptually but harder to reproduce, compare, and validate under realistic conditions.

Within this context, Scientific Large Scale Infrastructure for Computing/Communication Experimental Studies (SLICES) provides an enabling environment through its Cloud Continuum blueprint~{\cite{slicesds}}. The blueprint supports the deployment and observation of distributed applications across multiple continuum layers, allowing researchers to evaluate architectural choices, control logic, data-processing strategies, and placement decisions on realistic infrastructures rather than on purely simulated or ad hoc setups. HPC resources can therefore be considered as the computational back-end of the continuum, while Edge and Fog resources remain responsible for sensing, near-source coordination, and low-latency decisions.

This work is motivated by the five literature strands reviewed in Section~\ref{sec:related-work}. Existing studies provide solid foundations for research infrastructures, continuum architectures, scientific workflows, Digital Twins, and domain-specific applications, but these perspectives usually remain separated. Table~\ref{tab:related-positioning} summarizes the resulting gaps and shows how the proposed work addresses them through a workflow-oriented validation method that jointly considers infrastructure provisioning, application placement, timing, provenance, and domain semantics.

Building on this motivation, we discuss a generalized architecture for Cloud Continuum experimentation organized around three interacting layers: Edge, Fog, and Cloud~{\cite{computingcontinuumworkflows}}. The approach is not tied to a single domain; instead, it defines a reusable pattern in which lower layers host sensing and time-critical behavior, intermediate layers support near-source coordination and stream processing, and upper layers provide global analytics, persistence, and experiment-wide observability.

The approach is illustrated through two representative use cases: Renewable Energy Community (REC) management, focused on distributed Digital Twin coordination and community-level optimization, and AirWatch, a monitoring-oriented pipeline for low-latency alerting, fog-side anomaly detection, and cloud-side aggregation. These use cases show how the same substrate can support both control-oriented and monitoring-oriented workloads.

The contribution of this work is threefold. First, it defines a two-level reference architecture that separates the SLICES-based research-infrastructure layer from the application-level Edge--Fog--Cloud workflow organization. Second, it introduces a workflow-level methodology for instantiating continuum experiments by binding SLICES-managed resources with locally managed Edge and Fog resources. Third, it validates the approach through 40 automated runs over REC and AirWatch, showing how stage-level evidence can be collected and interpreted across virtualized and physical edge deployments.

\begin{table*}[!t]
\caption{Positioning of the proposed work with respect to the main literature strands.}
\label{tab:related-positioning}
\centering
\scriptsize
\begin{tabular}{p{0.18\textwidth}p{0.23\textwidth}p{0.23\textwidth}p{0.26\textwidth}}
\hline
\textbf{Literature strand} & \textbf{Main focus} & \textbf{Typical limitation for this paper} & \textbf{Role in our contribution} \\
\hline
Experimental infrastructures & Federation, programmability, resource access, and reproducible deployment substrates. & The experimental unit is often a slice, service, or platform configuration rather than a complete continuum workflow. & We use the SLICES Cloud Continuum Blueprint as the substrate over which workflow-level evidence can be generated and compared. \\
\hline
Continuum architectures & Placement, orchestration, latency, proximity, and management across Edge, Fog, and Cloud. & Architectural principles are not always translated into replayable descriptors and comparable validation procedures. & We make placement decisions explicit and evaluate them through two workflow profiles with different semantics. \\
\hline
Scientific workflows in the continuum & Composition, mapping, execution, and provenance of workflow stages across distributed resources; WMS orchestration across cloud and HPC. & WMS describe and execute workflows but assume the substrate as given; they do not provision, observe, and compare the underlying distributed infrastructure across runs. & Our experiment descriptors act as the infrastructure-level counterpart of a workflow description, providing resource bindings, provenance anchors, and stage-level evidence a continuum-aware WMS could consume. \\
\hline
Digital Twin models & Synchronization between physical and virtual entities, twin lifecycle, and cyber--physical correspondence. & The distribution of twin fragments across Edge, Fog, Cloud, and HPC resources is commonly left at a conceptual level. & We decompose Digital Twin responsibilities according to latency, aggregation, and optimization requirements. \\
\hline
Domain applications & Smart-energy control, community coordination, environmental sensing, and fog-enabled analytics. & Studies usually optimize one domain stack and one class of operational objective. & We compare REC and AirWatch as representative control-oriented and monitoring-oriented workflows on the same substrate. \\
\hline
\end{tabular}
\vspace{-0.4cm}
\end{table*}
Beyond these contributions, the paper emphasizes workflow observability as a core experimental requirement. In a continuum setting, correctness depends not only on whether a service produces an output, but also on where data are
generated, which services are handled by them, how long each transition is required, and whether the final decision or aggregation remains traceable to the corresponding edge-side observation.

The remainder of the paper is organized as follows. Section~\ref{sec:related-work} reviews the related literature and positions the proposed work with respect to the five strands summarized in Table~\ref{tab:related-positioning}. Section~\ref{sec:reference-architecture} introduces the two-level reference architecture and its Edge--Fog--Cloud organization. Section~\ref{sec:from-ref-to-exp} describes how the reference architecture is instantiated as executable continuum experiments. Section~\ref{sec:experimental-setup} presents the experimental setup, workloads, and deployment configurations, while Section~\ref{sec:evaluation} discusses the results obtained for REC and AirWatch. Finally, the paper concludes by summarizing the main findings and outlining future extensions.
\section{Related Work}
\label{sec:related-work}

The literature most relevant to this paper spans five partially overlapping strands: experimental research infrastructures, architectural models of the Cloud--Edge continuum, Digital Twin conceptualizations for CPS, and domain-oriented applications in smart energy and environmental monitoring. Reviewing these strands jointly is important because our contribution is not only an application design, but also a proposal for how heterogeneous continuum applications can be realized and comparatively validated on a research infrastructure.

\subsection{Platforms for Research Infrastructures}

Shared experimental infrastructures have long been recognized as essential to reproducible distributed-systems research. PlanetLab established the feasibility and scientific value of a programmable, wide-area facility for experimentation over a common substrate~\cite{planetlab}. GENI extended this perspective by emphasizing federation, network programmability, and at-scale experimentation across heterogeneous sites~\cite{geni}. In the IoT domain, FIT IoT-LAB showed that large-scale open experimentation can also be offered for constrained devices and wireless sensing scenarios, providing a reproducible environment for protocol and application validation~\cite{iotlab}. More recent Cloud-oriented facilities such as Chameleon demonstrate how testbeds can support interactive, multi-tenant, and artifact-aware experimentation for systems research~\cite{chameleon}. EdgeNet pushes the same logic toward multi-tenant Edge-Cloud experimentation, explicitly addressing scenarios that span beyond traditional datacenter boundaries~\cite{edgenet}. At a broader European research-infrastructure level, SLICES formalizes the need for a large-scale and integrated experimental facility for computing and communication studies~\cite{slicesds}. In particular, SLICES is especially relevant to Cloud Continuum research because it relies on federated and geographically distributed datacenter resources over which, the experimenters, can define dedicated experimental settings through specific blueprints, including those targeting computing-continuum scenarios.

These platforms are foundational for our work because they make controlled experimentation possible. However, many of them primarily address infrastructure provisioning, programmability, and federation at a general level. By contrast, the SLICES perspective is particularly suitable for Cloud Continuum experimentation because the federation of distributed sites can be specialized through blueprint-driven configurations that expose continuum-oriented environments instead of generic testbed slices only. This makes it possible to instantiate and compare different classes of Cloud Continuum applications (i.e., control-intensive and monitoring-intensive workflows) over a common and explicitly configured experimental substrate.

\subsection{Cloud Continuum Architectures}

The second strand concerns the architectural principles of computing-continuum. Early Fog computing work by Bonomi \textit{et al.} articulated the need to extend Cloud capabilities toward the network edge for latency-sensitive and geographically distributed IoT applications~\cite{fogrole}. Osanaiye \textit{et al.} discuss the transition from centralized Cloud systems toward Fog-oriented environments, emphasizing virtualization and live-migration challenges in distributed execution contexts~\cite{cloudtofog}. Chiang and Zhang frame Fog and IoT as a research space spanning networking, storage, control, and service placement across the Cloud-to-Things path~\cite{fogiot}. In parallel, Shi \textit{et al.} and Satyanarayanan provide two widely cited formulations of Edge computing, highlighting the importance of proximity, responsiveness, privacy, and resilience when data and computation move closer to users and devices~\cite{edgevision,edgeemergence}. Bittencourt \textit{et al.} further systematize the IoT--Fog--Cloud Continuum as an integrated problem involving infrastructure, management, and application concerns~\cite{iotfogcloud}. At the management level, Hong and Varghese survey resource-management mechanisms across Fog and Edge settings, underlining the role of heterogeneous resources, dynamic placement, and orchestration strategies~\cite{fogresource}. Svorobej \textit{et al.} complement this view by discussing orchestration from Cloud to Edge as a core requirement of the continuum rather than an implementation detail~\cite{orchestrationedge}.

Taken together, these studies define the architectural and management space to which this work belongs.
Nevertheless, they predominantly address what the continuum is and how it may be orchestrated, rather than how a reusable experimental blueprint should be structured.

\subsection{{Scientific Workflows in the Compute Continuum}}

{
The third strand concerns the execution and management of scientific workflows across the continuum. The literature on computing-continuum increasingly recognizes the need to coordinate resources across Edge, Fog, Cloud, and HPC environments, but less attention is often given to the experimental unit that should be reproduced and compared. A service-oriented evaluation may show that an individual microservice works correctly, while a platform-oriented evaluation may show that resources can be provisioned. Scientific workflows in the computing-continuum require a third viewpoint: the reproducible execution of a chain of stages whose meaning depends on placement, data movement, timing, and provenance.

This perspective builds on scientific-workflow research, where workflows are treated as structured executions involving: composition, mapping, execution, and provenance across distributed resources~\cite{workflowescience,provenanceworkflows}. It is also aligned with recent continuum-oriented works that frames Edge-to-Cloud applications as data-driven workflows and emphasizes the need to reproduce application behavior together with the physical and virtual infrastructure settings~\cite{computingcontinuumworkflows,e2clab}.
}

{
Workflow management systems make this viewpoint operational at the
application level. Established engines such as Pegasus automate the
mapping and execution of scientific workflows over distributed
resources~\cite{pegasus}, while continuum-oriented systems such as
StreamFlow explicitly support the hybrid execution of workflow steps
across Cloud and HPC environments~\cite{streamflow}. These systems
address how a workflow is \emph{described and executed}. The present
work addresses the complementary problem of how the distributed
substrate underneath such executions is \emph{provisioned, observed,
and made comparable} across runs. The experiment descriptors
introduced in Section~\ref{sec:descriptors}
can therefore be regarded as the infrastructure-level counterpart of a
workflow description: they do not replace a workflow management
system, but provide the resource bindings, provenance anchors, and
stage-level evidence that a continuum-aware system could consume.
}

{
This distinction is important because continuous workflows can fail in ways that are not detected by isolated component tests. A pipeline can preserve throughput but lose causal traceability; a control loop can produce correct decisions but out for the intended time window; an aggregation service can exhibit deterministic behavior while hiding upstream resource contention at the device level. 
Our evaluation therefore complements prior architectural work by asking whether the same experimental substrate can comparably expose these cross-layer effects. In this sense, the contribution is not only an application architecture, but also a method for making heterogeneous workflow behavior observable and reusable.
}

\subsection{Digital Twin Perspectives for CPS}

A fourth strand is the rapidly growing literature on Digital Twins. Kritzinger \textit{et al.} distinguish among Digital Model, Digital Shadow, and Digital Twin, providing a useful taxonomy for reasoning about the degree of synchronization between physical and virtual entities~\cite{dtmanufacturing}. Jones \textit{et al.} consolidate the field through a systematic review that highlights the need for clearer terminology, explicit twinning processes, and better characterization of the bidirectional links between physical and virtual systems~\cite{dtcharacterising}. Fuller \textit{et al.} broaden the discussion by surveying enabling technologies, open challenges, and application areas for Digital Twin systems~\cite{dtenabling}. These works are valuable because they move the discussion beyond a purely metaphorical use of the term ``Digital Twin'' and identify the technical requirements for maintaining coherent Cyber--Physical correspondence.

However, the majority of this literature remains centered on manufacturing or generic conceptual models. It does not fully address how a distributed Digital Twin should be partitioned across Edge, Fog, and Cloud layers in order to support time-sensitive coordination, human-in-the-loop decision-making, and repeatable experimentation on geographically distributed infrastructures. This is precisely the gap that our reference architecture targets.

\subsection{Application-Oriented Cloud Continuum Studies}

The fifth strand concerns concrete application domains. In the smart-energy area, Oprea and B{\^a}ra propose an Edge--Fog--Cloud architecture for IoT-based smart metering and show how multilayer processing can support distributed energy data management~\cite{smartmeterarch}. Their subsequent work on citizen energy communities emphasizes direct load optimization and flexibility services through Edge and Fog coordination~\cite{energycommunities}. Cicceri \textit{et al.} further present a learning-driven distributed Cyber--Physical architecture for RECs, combining Edge-to-Cloud execution with energy-aware adaptation~\cite{cicceri2023rec}. Beyond system architecture, Lowitzsch \textit{et al.} discuss RECs as an emerging governance and organizational model under the European Clean Energy Package~\cite{recpackage}, while Sousa \textit{et al.} survey peer-to-peer and community-based electricity markets, clarifying the market and coordination mechanisms that increasingly shape energy-community operation~\cite{p2pmarkets}. These studies confirm that energy communities are not only a control problem, but also a socio-technical coordination problem requiring distributed intelligence and adaptive resource management.

For urban-intelligence platforms, Tricomi \textit{et al.} propose an OpenStack-based architecture that integrates IoT management with the deployment of distributed services and edge-side workflows~\cite{tricomi2024urban}. In the environmental-monitoring area, Morawska \textit{et al.} review the maturity and limitations of low-cost sensing technologies for air-quality monitoring~\cite{airqualitysensing}. Castell \textit{et al.} experimentally assess commercial low-cost platforms and show both their promise and their data-quality limitations in practical deployments~\cite{airqualitycommercial}. Maag \textit{et al.} further systematize the calibration problem in air-pollution sensing networks, which is central when moving from isolated sensing devices to reliable distributed monitoring systems~\cite{airqualitycalibration}. Bharathi \textit{et al.} then exemplify how Fog enabled analytics can be used for air-quality monitoring and prediction close to the data source~\cite{airqualityfog}. These works collectively show that monitoring-oriented continuum applications must balance timeliness, data quality, and distributed processing.

Overall, the application literature validates the practical relevance of continuum solutions, but most studies optimize a single domain stack or a single operational concern. By contrast, our paper deliberately uses two distinct classes of applications, REC management and AirWatch-style urban monitoring, as representative workloads for evaluating a more general experimental approach to the Cloud Continuum. The key novelty is therefore not only in the individual application logic, but in the reusable architecture and experimental methodology that allow these applications to be realized and studied within the same infrastructure framework.
\section{Reference Architecture}
\label{sec:reference-architecture}

The reference architecture adopted in this paper 
is organized around two distinct but tightly connected layers. 
The first level concerns the research infrastructure that makes Cloud Continuum experimentation possible. The second level concerns the application architecture that can be instantiated on top of that infrastructure for concrete Cloud Continuum scenarios. This distinction is important because the architectural substrate used to expose, provision, and control the distributed resources is not the same thing as the application logic deployed over those resources. The current work therefore separates the enabling infrastructure architecture from the Edge--Fog--Cloud Continuum deployed above it~{\cite{slicesds}}.

\begin{figure*}
    \vspace{-0.4cm}
    \centering
\includegraphics[width=0.9\textwidth]{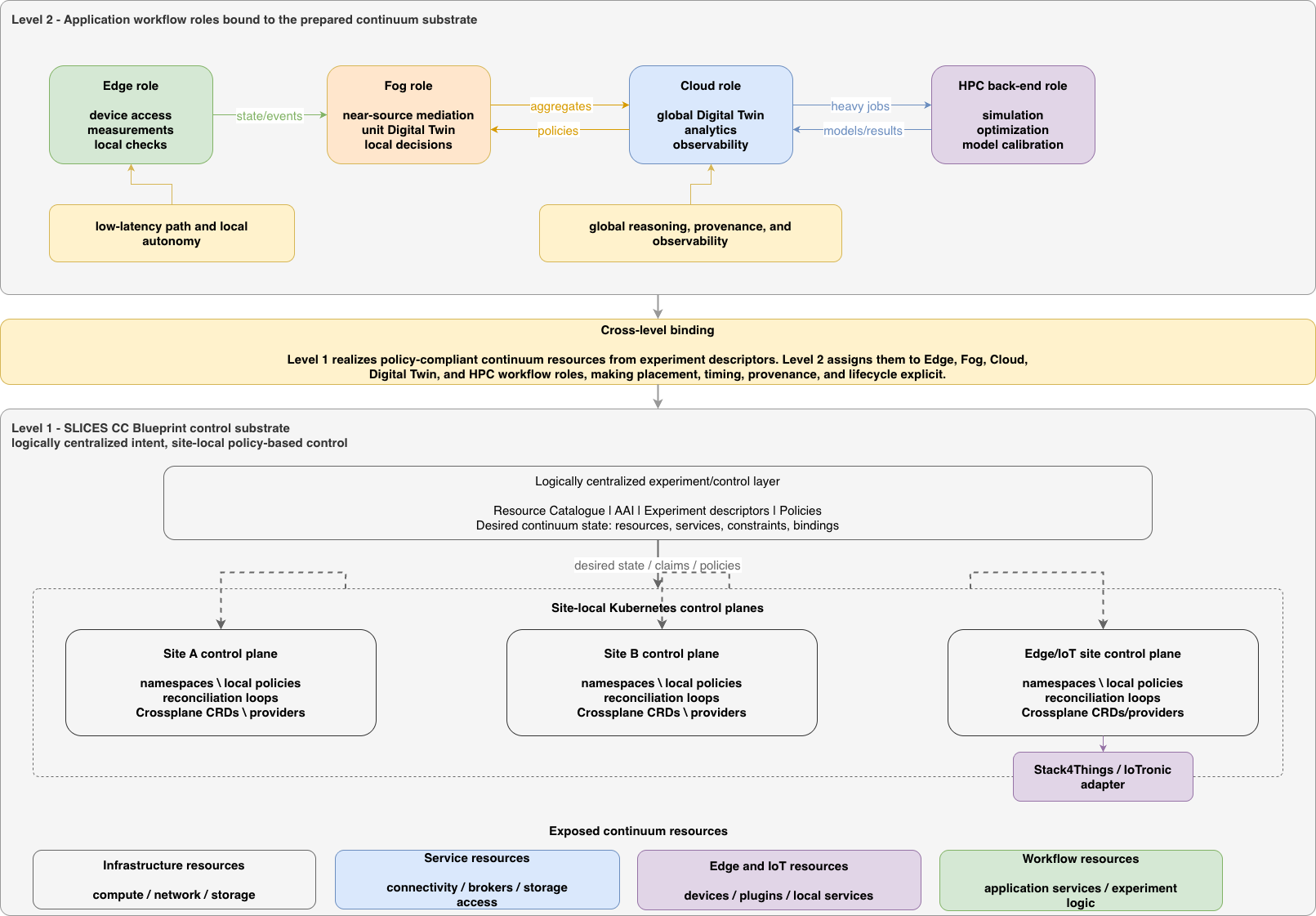}
\caption{Two-level reference architecture. The first level exposes and controls SLICES resources through the CC Blueprint, while the second level maps those resources to Edge, Fog, Cloud, Digital Twin, and HPC roles for Cloud Continuum applications.}
    \label{fig:two-level-reference-architecture}
    \vspace{-0.4cm}
\end{figure*}

Figure~\ref{fig:two-level-reference-architecture} summarizes this separation by showing the SLICES CC Blueprint as the infrastructure-level substrate and the Edge--Fog--Cloud Continuum as the application-level organization deployed on top of it.

\subsection{{Design Requirements for Continuum Experimentation}}

{
The two-level architecture is derived from a set of requirements that are specific to experimental Cloud Continuum research. These requirements are stricter than those of a conventional distributed application because the infrastructure must support both execution and scientific interpretation. A deployment that only runs successfully is not sufficient: the experimenter must also be able to explain the placement of each component, reconstruct the data path followed by each event, and compare alternative configurations without rebuilding the entire environment from scratch. The reference architecture therefore treats deployment, observation, and replay as first-class design objectives.
}

\begin{table*}[!t]
\caption{{Design requirements addressed by the proposed two-level architecture.}}
\label{tab:design-requirements}
\centering
\begingroup\scriptsize
\begin{tabular}{p{0.19\textwidth}p{0.35\textwidth}p{0.35\textwidth}}
\hline
\textbf{Requirement} & \textbf{Motivation} & \textbf{Architectural implication} \\
\hline
Explicit resource binding & Continuum experiments must state which physical, virtual, and federated resources are used in each run. & The infrastructure level exposes resources through catalogues, namespaces, and controlled provisioning mechanisms. \\
\hline
Layer-aware placement & Application functions have different latency, data-volume, and autonomy requirements. & The application level maps sensing, mediation, aggregation, optimization, and visualization to Edge, Fog, Cloud, and HPC roles. \\
\hline
Workflow traceability & Results must be attributable to the edge-side events and intermediate stages that generated them. & Messages, control decisions, and aggregates are interpreted as stages of a single workflow rather than isolated service outputs. \\
\hline
Controlled variability & Researchers need to vary devices, timing, placement, and workload pressure without losing comparability. & Experiment descriptors separate logical workflow definitions from concrete run-time bindings. \\
\hline
Multi-timescale execution & Cyber--physical applications combine immediate reactions, near-source coordination, global analytics, and long-running optimization. & Fast paths remain close to Edge and Fog resources, while Cloud and HPC resources support historical reasoning and compute-intensive analysis. \\
\hline
\end{tabular}
\endgroup
\vspace{-0.4cm}
\end{table*}

Table~\ref{tab:design-requirements} makes explicit the assumptions used to structure the architecture. The first requirement is explicit resource binding. In an ad hoc deployment, it may be sufficient to know that a container or device was available at execution time. In a research infrastructure, however, both the identity of the resource and its assigned role must be part of the experimental evidence. The second requirement is layer-aware placement. A Cloud Continuum workflow is not merely distributed; each function is placed according to a functional rationale. Sensing and safety checks should remain close to the physical process, coordination should be sufficiently near to bound latency and reduce upstream traffic, and global analytics should exploit higher-capacity resources. The third requirement is workflow traceability: the value of a result depends on the ability to link it to the observations and intermediate transformations that produced it.

The remaining requirements concern controlled variability and multi-timescale execution. Controlled variability is essential because the same workflow must be evaluated across different edge-device types, sampling rates, aggregation windows, and placement choices. These variations must be described explicitly; otherwise, different runs become difficult to compare. Multi-timescale execution is equally important because continuum applications rarely operate at a single temporal granularity. For example, a REC workflow may react to a local event, update a Fog-level Digital Twin fragment, and later trigger a Cloud- or HPC-based optimization stage. An AirWatch workflow may instead raise a near-real-time alert while also contributing to long-term environmental statistics. The architecture is therefore designed to connect these timescales without forcing all processing into a single layer.

\subsection{Level 1: Research-Infrastructure Architecture}

At the first level, the SLICES infrastructure enables the deployment of the CC Blueprint, which provides the architectural substrate used to expose geographically distributed resources as a coherent research infrastructure. The blueprint is designed as a modular environment for Cloud-native experimentation and distinguishes among three categories of resources: \emph{infrastructure resources}, \emph{service resources}, and \emph{workflow or application resources}. Infrastructure resources include compute, storage, and network assets, whether physical or virtual. Service resources provide the system-level capabilities needed to support an experiment, such as inter-cluster connectivity, storage access, or communication middleware. Workflow resources correspond to the experimental application that the researcher wants to deploy and study. This separation of concerns is essential because it allows experimenters to compose distributed environments without conflating the physical resources lifecycle, enabling services, and application-specific logic.

From an implementation perspective, the blueprint architecture  
is centered on a Kubernetes-based control model. Each participating site hosts a Kubernetes (K8s) cluster (at least an all-in-one node) that acts as the local execution and orchestration domain. Within each site, Kubernetes exposes a uniform control plane for deploying workloads, managing namespaces, and observing the state of the resources involved in the experiment. This choice makes the infrastructure Cloud-native by construction and allows the same operational abstractions to be reused across heterogeneous sites, even when the underlying hardware differs significantly.

On top of Kubernetes, the blueprint uses Crossplane to extend the control plane beyond in-cluster resources~{\cite{crossplane}}. Crossplane provides a declarative mechanism for provisioning and managing external resources through Kubernetes-native abstractions. In the CC Blueprint context, this means that distributed resources can be represented, configured, and managed through a single operational model rather than through site-specific ad hoc procedures. The Crossplane role therefore extends beyond deployment, providing a unifying mechanism through which the research infrastructure can be programmatically defined, managed, and reproduced.

At the infrastructure level, the blueprint also introduces central and site-local coordination components. A central controller maintains the \emph{Resource Catalogue}, which acts as the registry of the available distributed resources and of the configuration information needed to identify and use them across the infrastructure. Alongside it, an Authentication and Authorization service governs experiment access and token distribution across sites. At site level, local infrastructure providers retrieve the valid experiment configuration from the catalogue, prepare the execution environment, and configure the namespace in which the experiment will run. This step is especially important because it decouples the exposure of a resource from its use within a specific experiment: resources can be published, discovered, and prepared before the application workload is actually deployed.

The blueprint extends the set of available resources to support IoT and edge-device experiments. In particular, the integration with IoTronic, the Stack4Things Cloud component, makes it possible to expose IoT boards and remote devices as managed resources within the same control framework~{\cite{stack4things_longo2014}}.
Within the IoT module, each node is represented as a namespace-scoped custom resource, allowing device discovery and lifecycle management to remain aligned with the surrounding experimental environment~{\cite{dagati2025iotorchestration}}. Namespace-registration mechanisms can then deploy plugins, services, and Function-as-a-Service (FaaS)-based functions directly to registered devices while preserving experiment-level isolation and control~{\cite{faasiot}}. Consequently, the infrastructure is not restricted to Cloud-hosted workloads, but incorporates physical edge-nodes, remote sensors, and device-side computation as first-class elements of the same research substrate.

From an operational viewpoint, the first level enables the experimenter to publish a distributed resource description into the catalogue: after authentication through the SLICES access services, the experimenter can trigger the preparation of execution domains across the selected sites.
Site-local providers configure namespaces and retrieve the resources relevant to the experiment. Once the environment is ready, an experiment-level controller can deploy the distributed workflow across multiple clusters and, when needed, inject plugins or services into IoT nodes. In this way, the CC Blueprint provides the infrastructure-level capabilities required for repeatability, federation, multi-site deployment, and uniform lifecycle management.

\subsection{Level 2: Application Architecture on Top of the Blueprint}

The second level of the reference architecture concerns the realization of Cloud Continuum applications over the infrastructure previously described. At this level, the focus shifts from how the research infrastructure exposes and manages resources to how those resources are organized into an application architecture. In this paper, this second level is modeled as an Edge--Fog--Cloud architecture for distributed Cyber--Physical applications. The purpose of this level is not to redefine the underlying infrastructure, but to define how application responsibilities are partitioned across the resources made available upon the blueprint level.

The Edge layer hosts the functions that must remain closest to the physical process. These include device interfacing, raw data acquisition, lightweight preprocessing, local buffering, and emergency actions that cannot depend on round trips to upper layers. When the blueprint is extended with IoT support, edge nodes can be represented and managed as infrastructure-level resources while still executing application-specific logic at this second level. 

The Fog layer hosts the intermediate logic that benefits from proximity to the physical environment while still requiring more flexibility than the Edge. This is the layer in which local controllers, protocol mediators, stream-processing services, anomaly-detection functions, queue managers, or unit-level Digital Twin components can be deployed. In the infrastructure view, these services are simply workloads deployed inside the namespaces prepared by the blueprint. In the application view, however, they represent an operational mediation point between the immediacy of device-side behavior and the global reasoning performed at Cloud level.

The Cloud layer hosts the components that benefit from a global view of the experiment and from larger pools of compute and storage resources. This includes persistent repositories, model execution, experiment-wide analytics, visualization services, optimization modules, and long-term observability functions. Again, the distinction between the two levels remains important: at the infrastructure level, 
resources (taken from multi-site Clouds) are 
configured as distributed compute and storage assets under a common control plane; at the application level, they become the place where global knowledge is consolidated and where cross-site coordination logic is executed, 
in an experimental form.

When the application is based on a distributed Digital Twin, the same Edge--Fog--Cloud partitioning defines how the twin is decomposed across the continuum. Edge nodes maintain the closest representation of the physical process by acquiring measurements, enforcing local checks, and updating a local shadow of the observed device or environment. Fog nodes host unit-level Digital Twin fragments, 
which aim to combine 
recent measurements, local state, user or operator requests, and coordination policies in order to support near-source decision-making. Cloud nodes maintain the global Digital Twin view, where data from multiple units are consolidated and used for analytics, prediction, optimization, and experiment-wide reasoning. This decomposition makes the Digital Twin operationally compatible with latency constraints, local autonomy, and global coordination requirements.

The Cloud layer may also include, or interoperate with, HPC resources when the application requires computationally intensive back-end functions. In this role, HPC resources complement the continuum by supporting large-scale simulation, optimization, model calibration, uncertainty analysis, and historical analysis, while Edge and Fog components continue to handle sensing, local control, and low-latency interactions. This motivates the execution of experiments specifically aimed at assessing HPC-oriented techniques within Cloud Continuum applications: the objective is not simply to move computation to a more powerful back-end, but to determine which application functions can benefit from parallel execution, accelerated numerical kernels, large-scale data processing, or batch-oriented simulation without breaking the latency, locality, and autonomy constraints of the Edge--Fog--Cloud workflow. Such experiments make it possible to quantify the trade-offs among execution time, data movement, orchestration overhead, scalability, produced models' accuracy, and Cyber--Physical control loop responsiveness. The architecture therefore keeps real-time Cyber--Physical behaviour close to the physical process, but still allows heavier analytical workloads to be offloaded to specialized Cloud or HPC back-ends when this improves the overall application execution.

{
An important consequence of this organization is that the architecture separates fast operational paths from heavier analytical paths without treating them as independent systems. Edge and Fog components preserve responsiveness and local autonomy, while Cloud and HPC components provide broader context, historical state, and computational capacity. The same event can therefore participate in multiple temporal regimes: it may trigger an immediate edge-side or fog-side reaction, contribute to a cloud-side aggregate, and later feed an optimization or simulation stage. This multi-timescale interpretation is central to continuum workflows because it reconciles low-latency control with long-running scientific analysis.
}

{
The separation between infrastructure and application levels also supports portability. A given workflow stage can be re-bound to a different site, resource type, or execution tier while preserving the logical application structure. For example, anomaly detection may be evaluated as a fog-side function in one deployment and as an edge-side function in another; similarly, optimization may be executed as a Cloud service or offloaded to an HPC back-end. The reference architecture therefore provides a vocabulary for describing these changes as controlled placement variations rather than as unrelated implementations.
}

The key contribution of the two-level formulation is therefore the explicit mapping between infrastructure capabilities and application roles. The first level makes available a federated, programmable, and repeatable experimentation substrate; the second level uses that substrate to realize a concrete Cloud Continuum architecture. This means that the same infrastructure can host very different application classes. A control-oriented application may exploit the second level to implement time-window-based coordination, local scheduling, and predictive optimization. A monitoring-oriented application may instead instantiate validation pipelines, anomaly detection, Cloud aggregation, and dashboards.

\subsection{Cross-Level Mapping}

The relationship between the two levels can be summarized through a simple mapping principle. The infrastructure level determines \emph{where} and \emph{how} resources are exposed, authenticated, provisioned, and controlled. The application level determines \emph{which role} those resources play within a Cloud Continuum workflow. A Kubernetes site prepared by the blueprint may host cloud-side analytics in one experiment and fog-side coordination services in another. An IoT board integrated through Stack4Things may serve as a sensing node in a monitoring pipeline or as a controllable actuator in a smart-energy scenario. This separation keeps the architecture reusable while preserving enough structure to make experiments comparable and repeatable.

Within this perspective, the architecture described in the remainder of the paper should be interpreted as a second-level architecture instantiated over the first-level SLICES substrate. REC management and AirWatch are representative application profiles realized on top of the same research-infrastructure blueprint. The value of the proposal lies mainly in this separation: the CC Blueprint provides the infrastructure-level foundation for experimentation, while the Edge--Fog--Cloud architecture provides the application-level pattern through which Cloud Continuum solutions can be designed, deployed, and validated.
\section{From Reference Architecture to Experimental Scenarios}
\label{sec:from-ref-to-exp}

The SLICES CC Blueprint provides a high-level, generic methodology for federating distributed computing resources. However, translating this theoretical framework into a reproducible experimental testbed requires a structured instantiation process. This section details how the abstract SLICES primitives are mapped onto a concrete Edge--Fog--Cloud deployment topology by introducing the specific application profiles used to validate the architecture in a realistic setting.

\subsection{Instantiation Process}
The transition from the reference CC Blueprint to a functional experimental scenario involves defining the precise role and placement of resources across the continuum. Rather than treating the infrastructure as a flat pool of resources, the instantiation process organizes the deployment into three distinct functional tiers. 

At the outermost boundary, Edge provisioning involves configuring resource-constrained physical devices, such as Raspberry Pi units, or emulated containerized environments (Pods) to act as the primary data ingress points. Here, the edge-side nodes are equipped with lightweight daemon processes responsible for handling local sensor interfacing, initial data normalization, and publishing event streams via secure tunnels to higher tiers. Moving inward, Fog delegation allocates intermediate nodes (which are geographically or topologically positioned between the Edge and the Cloud) to host message brokers and localized analytics. This intermediate processing is crucial for bounding latency and reducing the raw data volume transmitted upstream. Finally, Cloud centralization dedicates high-capacity data center resources to host heavy, stateful aggregation frameworks and global decision-making logic. By formalizing this three-tier mapping, the SLICES infrastructure is effectively transformed from a raw hardware pool into a structured Cyber-Physical experimentation substrate.

\begin{figure*}[!t]
    \vspace{-0.4cm}
    \centering
    \resizebox{0.85\textwidth}{!}{\input{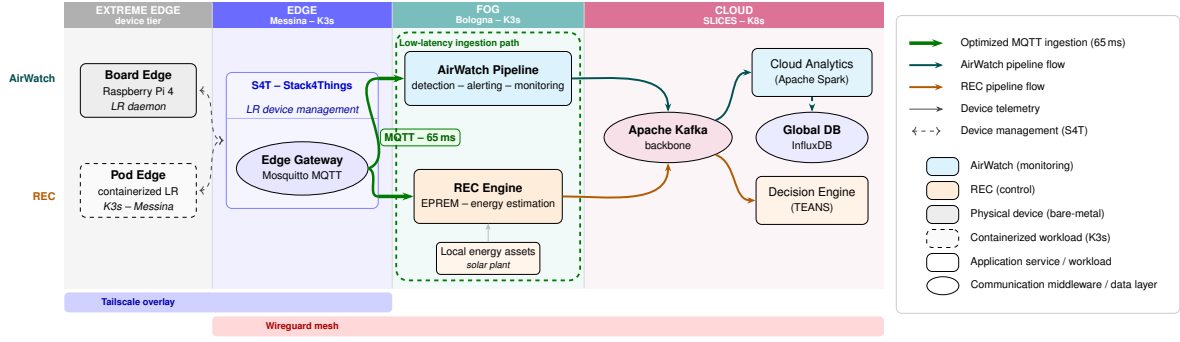}}
    \caption{End-to-end dataflow of the instantiated Edge--Fog--Cloud topology.
    The Extreme Edge tier hosts physical Raspberry Pi boards or containerized
    pods running the Lightning-rod daemon. The Edge Gateway in Messina exposes
    an MQTT broker as the low-latency ingestion point ($\sim$65~ms).
    The Fog tier in Bologna hosts the AirWatch pipeline and the REC Engine,
    which forward-process streams via Apache Kafka toward cloud-side analytics
    (Spark) and the TEANS\cite{cicceri2023rec} decision engine. Device management is handled by
    Stack4Things across both edge configurations.}
    \label{fig:dataflow}
    \vspace{-0.4cm}
\end{figure*}

\subsection{Application Profiles}
To demonstrate the versatility of the architecture, we define two contrasting application profiles designed to represent fundamentally different classes of continuum workloads.

The first is the \textit{Control-Oriented Profile}, exemplified by the REC scenario. This profile focuses on closed-loop, latency-sensitive operations requiring the continuum to support distributed Digital Twin coordination. In this scenario, telemetry from Edge prosumers must be rapidly routed through a 
MQTT broker toward Fog processing logic for distributed Digital Twin coordination. The primary operational requirement here is deterministic, time-window-based control to maintain system equilibrium, such as dynamically scheduling flexible loads to preserve global energy balance.

In contrast, the \textit{Monitoring-Oriented Profile}, represented by the AirWatch scenario, focuses on high-throughput data pipelines and stateful analytics over time. It requires the Edge tier to continuously ingest complex, multidimensional environmental metrics. The continuum's role is to abstract the underlying hardware heterogeneity, allowing the Cloud to perform reliable window-based data reduction and anomaly detection via streaming analytics, ultimately providing long-term urban observability and alerting capabilities.

\subsection{Configurable Parameters and Validation Objectives}
A key advantage of deploying these diverse application profiles over the CC Blueprint is the ability to conduct controlled, reproducible experiments. The architecture exposes several configurable parameters that researchers can manipulate to stress-test the continuum. These include temporal parameters, such as sampling intervals at the Edge and aggregation windows at the Cloud, as well as structural choices regarding the placement of processing logic across the tiers. Furthermore, the setup is designed to accommodate infrastructure constraints (e.g., to simulate network fluctuations or node failures in future benchmarking studies).

By applying both the REC and AirWatch profiles to the instantiated topology, our experimental campaign aims to address specific validation objectives. Primarily, it seeks to provide evidence of architectural reusability by demonstrating that identical infrastructural primitives (such as the Lightning-Rod daemon, Apache Kafka, and Apache Spark) can host both control-oriented and monitoring-oriented workloads.
Additionally, the evaluation focuses on empirical performance profiling/measuring the system's ability to maintain low-latency event-driven ingestion and steady throughput across both ideal simulated environments and constrained physical setups. Ultimately, the goal is to verify the architecture's capacity to maintain end-to-end data provenance for crucial control decisions while ensuring deterministic data aggregation for continuous monitoring tasks.

The transition from reference architecture to experiment also defines the unit of comparison used throughout the evaluation. Each run is interpreted as an instance of the same continuum workflow pattern, but with different workload semantics and resource constraints. This makes it possible to compare not only raw throughput or latency, but also how the architecture preserves workflow properties such as traceability, deterministic aggregation, and recovery from heterogeneous edge behavior. In this sense, the experiment is not limited to validating a single implementation; it validates whether the same blueprint can support repeatable reasoning across different classes of scientific and Cyber--Physical workflows. The evaluation in Section~\ref{sec:evaluation} quantifies this delta across the two configurations.

To keep this comparison meaningful, the mapping between abstract roles and concrete resources must be explicit. Edge resources are associated with data production and local buffering; Fog resources are associated with near-source mediation and latency observation; Cloud and HPC resources are associated with global analytics, aggregation, optimization, and historical reasoning. Reporting these bindings clarifies where each result originates and allows future experiments to modify one placement decision at a time while preserving the rest of the workflow structure.

\subsection{Workflow Decomposition Across the Continuum}

The two application profiles are instantiated through the same decomposition logic. Each workflow is expressed as a sequence of stages, and each stage is associated with a continuum role rather than with a fixed machine. This is important because the experimental goal is not to prove that one particular node can execute one particular task, but to show that the same infrastructure can host, observe, and compare different allocations of the same logical stages. Table~\ref{tab:workflow-decomposition} summarizes this decomposition.

\begin{table*}[!t]
\caption{Workflow-stage decomposition used to instantiate REC and AirWatch over the same continuum substrate.}
\label{tab:workflow-decomposition}
\centering
\scriptsize
\begin{tabular}{L{0.17\textwidth}L{0.24\textwidth}L{0.24\textwidth}L{0.25\textwidth}}
\hline
\textbf{Continuum role} & \textbf{REC control workflow} & \textbf{AirWatch monitoring workflow} & \textbf{Experimental evidence collected} \\
\hline
Edge stage & Prosumer-side measurements, local status updates, and safety-oriented checks. & Sensor acquisition, local formatting, threshold checks, and emergency event generation. & Device type, sampling interval, timestamp of generation, local buffering behavior. \\
\hline
Fog stage & Unit-level Digital Twin fragment, message mediation, local coordination, and latency observation. & Stream validation, enrichment, anomaly pre-filtering, and alert forwarding. & Edge-to-Fog delay, message continuity, broker responsiveness, intermediate timestamps. \\
\hline
Cloud stage & Community-level Digital Twin view, global decision engine, historical storage, and dashboarding. & Window-based aggregation, time-series persistence, historical visualization, and reporting. & Aggregation determinism, end-to-end provenance, window completeness, decision or summary latency. \\
\hline
{HPC back-end stage}\tnote{*} & Optimization, simulation, scenario exploration, and model calibration for future scheduling windows. & Offline calibration, long-term trend analysis, and sensitivity studies over historical measurements. & Batch duration, data-transfer overhead, model output availability. \\
\hline
\end{tabular}
\begin{tablenotes}
\item[*] {
The HPC stage is fully specified in the decomposition, including the evidence it must produce (batch duration, data-transfer overhead, model output availability); its experimental exercise over the same run descriptors is the first target of the follow-up campaign.
}
\end{tablenotes}
\vspace{-0.4cm}
\end{table*}

The decomposition also provides a practical criterion for deciding whether an application function should be moved across layers. A stage is a candidate for Edge placement when it reduces unsafe delay, protects local autonomy, or prevents unnecessary upstream traffic. Moreover, it is again a candidate for Fog placement when it requires a broader local context, but still needs to remain close to the data source.  
Similarly, we select a stage for Cloud placement when it benefits from global state, persistent storage, and integration with dashboards or decision engines; and finally, it is chosen for HPC back-end or specialized back-end execution when the computation is too heavy for the online path, can tolerate batch-oriented scheduling, and produces outputs that improve future decisions rather than immediate reactions.

This stage-based view prevents the two use cases from being interpreted as unrelated prototypes. REC and AirWatch differ in semantics, but they share the same experimental grammar: edge-side generation, fog-side mediation, cloud-side consolidation, and back-end analytics. The comparison therefore focuses on how different semantics stress the same grammar. In REC, the critical question is: \textit{Can a measurement be traced back to a timely control decision?} In AirWatch, the critical question is: \textit{Can a continuous stream be transformed into stable alerts and aggregates?}. The common stage decomposition allows both questions to be studied without changing the underlying infrastructure model.

\subsection{Experimental Run Lifecycle}

We define an experimental run as an execution instance of an experimental template defined upon the research infrastructure. 
Each experimental run follows a lifecycle that connects the abstract architecture with concrete measurements, so that it is possible to describe it through a series of five steps. First, the experimenter selects the application profile and identifies the continuum roles that must be instantiated. Second, the resource binding is resolved through the infrastructure-level mechanisms: the selected Edge, Fog, Cloud, and HPC back-end resources are associated with the run, and the required namespaces, services, and communication paths are prepared. Third, the workload is initialized by configuring input generators, sampling intervals, thresholds, window sizes, and duration. Fourth, the run is executed while telemetry, application logs, and timestamps are collected. Finally, the collected evidence is interpreted according to the workload semantics, distinguishing between control timeliness in REC and monitoring stability in AirWatch.

This lifecycle is deliberately conservative. It avoids treating deployment automation as the only indicator of reproducibility and instead requires the run configuration to remain connected with the resulting measurements. For example, an increase in latency is meaningful only if the corresponding device type, placement, and workload pressure are known. Similarly, a successful aggregation window is meaningful only if the window boundaries, input rate, and message ordering assumptions are recorded. The lifecycle therefore binds together infrastructure preparation, application configuration, and evaluation semantics.

The same lifecycle can also support negative or stress-oriented experiments. A run can intentionally introduce a constrained edge-device, a higher sampling frequency, a delayed upstream path, or a different placement of analytical logic. Because the rest of the descriptor remains stable, the effect of the change can be interpreted as a controlled variation rather than as a separate experiment. This property is essential for progressively building a benchmark suite for cloud-continuum workflows.

\subsection{Experiment Descriptors and Replayability}
\label{sec:descriptors}

The experimental scenarios are also defined through descriptors that make each run reproducible. A descriptor record comprises the selected use case, deployment mode, resource binding, input generator, timing parameters, active services, and metrics to be collected. This descriptor-oriented approach is useful because it separates the workflow definition from a single execution instance. The same AirWatch or REC workflow can be repeated with a different edge-device, a different sampling interval, or a different aggregation window while preserving the logical structure of the experiment.

Reproducibility is especially important when synthetic domain values are used. Since CO$_2$ and energy traces are generated for controlled experimentation, the input configuration should be treated as part of the experiment. Recording the generator settings allows researchers to reproduce the same workload pressure on the continuum substrate. Furthermore, it enables the introduction of alternative traces at a subsequent stage without necessitating any alteration to the infrastructure-level interpretation of the results.

The descriptor further supports comparison between normal and stressed executions. For example, a run may be characterized by stable connectivity, while another may include delayed messages, bursty arrivals, or physical-device overhead. By preserving the descriptor, the evaluation can distinguish whether a change in latency or throughput is caused by workload semantics, device constraints, network behavior, or placement decisions. This is a necessary step toward building a reusable benchmark for cloud-continuum workflows rather than a one-time demonstration.

{
This descriptor-oriented organization is also intended to align the experimental methodology with the Findable, Accessible, Interoperable, Reusable (FAIR) principles. Run descriptors make each experiment findable and identifiable as a distinct, citable configuration; the public artifact repository, which includes the orchestration scripts, Kubernetes manifests, and raw datasets, makes the corresponding evidence accessible; the exclusive use of standard open-source components and declarative manifests supports interoperability across sites; and the separation between the logical workflow definition and its concrete run-time binding is precisely what makes both the data and the experimental procedure reusable by third parties. 
}
\section{Experimental Setup}
\label{sec:experimental-setup}

The experimental campaign validates the proposed continuum substrate across two workloads with different coordination requirements: the REC control workflow and the AirWatch monitoring pipeline. Both are deployed over the same Edge--Fog--Cloud infrastructure, instantiated through the SLICES Cloud Continuum blueprint, with Cloud services hosted on the SLICES infrastructure, Fog components in Bologna, and Edge components in Messina. This geographical distribution makes cross-layer interactions observable under realistic network conditions rather than in a controlled local environment.

\subsection{Deployment Topology and Software Stack}
\label{sec:deployment-topology}

The continuum is built on standard open-source components to ensure reproducibility and site independence. 
At the Edge, physical Raspberry Pi devices or simulated Kubernetes Pods run the Lightning-Rod daemon (Stack4Things edge agent), which handles local sensor ingestion and establishes a secure overlay connection toward the local gateway in Messina. In the physical configuration, IoTronic reachability is provided either through a public endpoint or through an overlay network.
This gateway hosts a broker (Eclipse Mosquitto) that serves as the entry point for low-latency data ingestion. Moving inward, the Fog tier in Bologna hosts custom processing modules (EPREM for REC, ingest for AirWatch) that perform near-source mediation, operating as event-driven subscribers to the Messina-based MQTT broker via an overlay network as a backbone (in this specific case, WireGuard). At the Cloud tier, Apache Spark Structured Streaming handles stateful window-based aggregation, while the TEANS engine executes global Digital Twin decisions for REC, employing Apache Kafka for inter-site synchronization and data propagation.

Two deployment configurations are compared throughout the evaluation. \textbf{POD} is a cloud-native baseline in which the edge function runs as a Kubernetes pod. \textbf{RASP} is a physical edge configuration in which the same function runs on a Raspberry Pi~4 Model~B (Cortex-A72 quad-core, 1.8~GHz, 4~GB LPDDR4). Both configurations share the same MQTT-based workflow logic and differ only in the Edge-tier realization, enabling a controlled comparison between virtualized and physical edge behavior. Table~\ref{tab:setup} summarizes the complete setup parameters.

\begin{table*}[!t]
\caption{Experimental setup summary.}
\label{tab:setup}
\centering
\small
\setlength{\arrayrulewidth}{0.6pt}
\renewcommand{\arraystretch}{1.18}
\begin{tabular}{|L{0.20\textwidth}|L{0.34\textwidth}|L{0.34\textwidth}|}
\hline
\textbf{Aspect} & \textbf{POD} & \textbf{RASP} \\
\hline
\celllines{Edge\\realization} & Kubernetes pod & \celllines{Raspberry Pi~4 Model~B\\Cortex-A72 quad-core 64-bit 1.8~GHz\\4~GB LPDDR4} \\
\hline
\celllines{Fog substrate} & \multicolumn{2}{|L{0.68\textwidth}|}{\celllines{QEMU VM / Stack4Things (S4T) node 16 cores 15.57~GB RAM}} \\
\hline
\celllines{Cloud substrate} & \multicolumn{2}{|L{0.68\textwidth}|}{SLICES shared infrastructure\tnote{*}} \\
\hline
\celllines{Network\\paths} & \multicolumn{2}{|L{0.68\textwidth}|}{\celllines{Board$\to$Messina Gateway via Tailscale (RASP configuration only)\\Messina$\to$Bologna Fog via WireGuard backbone\\Fog$\to$Cloud through the Bologna/SLICES WAN path}} \\
\hline
\celllines{Software stack} & \multicolumn{2}{|L{0.68\textwidth}|}{\celllines{Lightning-rod v0.4.7; Eclipse Mosquitto v2.1.2; Kafka v3.7.0; Spark v3.5.0; InfluxDB v1.8.10}} \\
\hline
\celllines{Experiment duration} & \multicolumn{2}{|L{0.68\textwidth}|}{\celllines{10-minute duration; 5~s sampling interval}} \\
\hline
\celllines{AirWatch windows} & \multicolumn{2}{|L{0.68\textwidth}|}{\celllines{60~s Spark tumbling windows over 9 environmental variables}} \\
\hline
\celllines{REC edge nodes} & \multicolumn{2}{|L{0.68\textwidth}|}{\celllines{Two physical-emulated nodes: \textit{bologna-plant-01} and \textit{messina-house-01} (MQTT stream)}} \\
\hline
\celllines{Payload size} & \multicolumn{2}{|L{0.68\textwidth}|}{\celllines{$\approx$300~B for REC; $\approx$600~B for AirWatch}} \\
\hline
\end{tabular}
\begin{tablenotes}
\item[*] Specific resource allocation is managed by the provider and not exposed to external experiments.
\end{tablenotes}
\vspace{-0.4cm}
\end{table*}

\subsection{Experiment Definition and Measurement Protocol}
\label{sec:instrumentation}

The deployment of experiments (considering both virtualized infrastructure and application to be studied) involves 
every tier of the Research Infrastructure. To ensure measurement consistency, all compute nodes across Messina, Bologna, and the SLICES Cloud sites are synchronized via the Network Time Protocol (NTP), with measured clock offsets typically under 1~ms. At the Edge, each observation is timestamped ($t_{gen}$) immediately upon acquisition and annotated with device type and provenance identifier. At the Fog, the protocol records the arrival time ($t_{arr}$) at the MQTT ingestion service ($s_1$). Ingestion latency is calculated as $L_{ing} = t_{arr} - t_{gen}$, accounting for the multi-hop network transit. At the Cloud, Spark is used to log the timestamps ($t_{comp}$) when an aggregation window is finalized. Processing latency is reported as $L_{proc} = t_{comp} - t_{window\_end}$, representing the duration between the logical close of a time window and the availability of its aggregate result.

The evaluation consists of a systematic campaign of 40 automated runs (10 independent repetitions per scenario and configuration) to ensure the stability of the reported observations and account for transient network variations across the distributed sites. Each run is characterized by a fixed descriptor recording device type, sampling interval, aggregation window, active services, and placement. Results are reported as means and standard deviations across the aggregated batches composed by a set of 10 repetitions. Distributions are analyzed using the Interquartile Range (IQR) method to identify and characterize transient network outliers (typically $<1\%$ of samples) separately from the steady-state pipeline performance.

{
\subsection{Continuum Orchestration and Reproducibility}
\label{sec:continuum-orchestration}
The transition from a high-level blueprint to a running experiment follows a structured orchestration process. Once access to the SLICES infrastructure is obtained, the user selects the required compute and network resources across the distributed sites. 
An instance of Lightweight Kubernetes
(K3s) is deployed on the allocated nodes to instantiate K3s clusters across the selected sites.\\
A centralized deployment orchestrator 
manages the software stack lifecycle. It automates the instantiation of the cloud-side analytics 
and messaging backbone, 
the fog-side mediation modules, and the edge-side ingestion services. Physical edge-devices, such as Raspberry Pis, are integrated into this continuum by attaching them to the \textit{Stack4Things} (S4T) substrate. This is achieved by running the S4T \textit{Lightning-Rod} daemon on the physical hardware, which establishes a secure overlay tunnel to the local gateway, effectively bridging the physical resources with the virtualized Kubernetes environment.\\
To ensure reproducibility, the entire campaign is managed by an automated runner 
that performs state resets and extraction of logs and timestamps across all sites. The complete research artifact, including the orchestration scripts, Kubernetes manifests, and raw datasets, is available in a dedicated repository at
\url{https://github.com/ProSoDiAC/cloud-continuum-cps-infra}.
}

\subsection{Evaluation Questions}
\label{sec:evaluation-questions}

The evaluation is organized around four questions. \textit{Q1}: Can the same continuum substrate host both a control-oriented and a monitoring-oriented workflow without changing the underlying infrastructure? \textit{Q2}: Does the architecture preserve per-stage evidence (provenance, timestamps, and stage-level observations) sufficient to interpret results after execution? \textit{Q3}: Does the physical edge configuration introduce measurable overheads with respect to the pod-based baseline? \textit{Q4}: Does cloud-side consolidation remain stable enough to produce comparable outputs across configurations?

These questions are intentionally broader than a single performance metric. The goal is to verify that the infrastructure exposes the evidence needed to reason about timing, placement, semantic correctness.
Threats to the validity of these answers include device heterogeneity between POD and RASP, network variability across distributed sites, and the use of synthetic workload generators.
These threats are mitigated by keeping the run descriptor explicit and by decomposing results per stage, so that any observed difference can be traced to a specific tier rather than attributed to the workflow as a whole.
\section{Experimental Evaluation}
\label{sec:evaluation}

This section presents the results of the experimental campaign conducted over the instantiated Edge--Fog--Cloud architecture. As stated above, the evaluation consists of 40 automated runs (10 independent repetitions per scenario and configuration). Reported values are: arithmetic means and standard deviations across these 10 repetitions unless otherwise noted. The evaluation follows the methodology and addresses the evaluation Questions (Q1--Q4) defined in Section~\ref{sec:experimental-setup}.

\subsection{Use Case 1: Renewable Energy Community}
The REC workflow evaluates coordination using a low-latency push path. Data is streamed from edge prosumers to the fog-side logic for real-time Digital Twin synchronization.

The POD and RASP configurations show comparable throughput, resulting in a matching number of TEANS decisions across runs. All runs produced consistent flexible-load scheduling outcomes, and global energy-balance distributions remained stable across configurations, suggesting that, under the tested traces, the control logic is not materially affected by the edge realization. The high variance in energy balance reflects the expected oscillation of the community around the equilibrium point; notably, these metrics are derived from synthetic energy traces designed to stress the balancing logic under controlled experimental conditions.

\begin{figure*}[t]
     \vspace{-0.5cm}
    \centering
    \resizebox{0.7\textwidth}{!}
    {
\pgfplotsset{
  paper box/.style={
    font=\footnotesize,
    grid=major, grid style={gray!20, very thin},
    axis line style={gray!60},
    tick style={gray!60},
    boxplot/draw direction=y,
    boxplot/box extend=0.32,
    cycle list name=black white,
  },
  aw box/.style={fill=cyan!15!white, draw=teal!70!black, thick,
                 mark=o, mark options={draw=teal!70!black,fill=white,scale=0.95}},
  rec box/.style={fill=orange!15!white, draw=orange!70!black, thick,
                  mark=o, mark options={draw=orange!70!black,fill=white,scale=0.95}},
}

\begin{tikzpicture}

\begin{axis}[
  paper box,
  name=panelA,
  width=5.0cm, height=6.5cm,
  title={\textbf{(a)} Ingestion latency},
  title style={font=\small\bfseries, at={(0.5,1.04)}},
  ylabel={Latency (ms)},
  ylabel style={font=\footnotesize},
  ymin=60.0, ymax=72.0,
  ytick={60,62,64,66,68,70},
  xtick={1,2},
  xticklabels={AirWatch, REC},
  xticklabel style={font=\footnotesize},
  xmin=0.4, xmax=2.6,
  ymajorgrids=true,
]
  \addplot[aw box, boxplot prepared={
    draw position=1,
    lower whisker=60.76,
    lower quartile=62.93,
    median=63.84,
    upper quartile=64.88,
    upper whisker=67.80,
  }] coordinates {};

  \addplot[rec box, boxplot prepared={
    draw position=2,
    lower whisker=60.32,
    lower quartile=63.35,
    median=65.01,
    upper quartile=66.11,
    upper whisker=70.25,
  }] coordinates {};

  \addplot[teal!60!black, dashed, thin, forget plot]
    coordinates {(0.55,64.49)(1.45,64.49)};
  \addplot[orange!60!black, dashed, thin, forget plot]
    coordinates {(1.55,65.27)(2.45,65.27)};

  \node[font=\scriptsize, text=teal!70!black, anchor=west]
    at (axis cs:1.48,64.49) {$\mu$=64.5};
  \node[font=\scriptsize, text=orange!70!black, anchor=west]
    at (axis cs:2.06,65.27) {$\mu$=65.3};
\end{axis}

\begin{axis}[
  paper box,
  name=panelB,
  at={($(panelA.east)+(1.4cm,0)$)}, anchor=west,
  width=9.5cm, height=6.5cm,
  title={\textbf{(b)} Pipeline stages (log scale)},
  title style={font=\small\bfseries, at={(0.5,1.04)}},
  ylabel={Latency (ms, log)},
  ylabel style={font=\footnotesize},
  ymode=log,
  ymin=0.2, ymax=8000,
  ytick={0.3,1,3,10,30,100,300,1000,3000},
  yticklabels={0.3,1,3,10,30,100,300,1\,000,3\,000},
  yticklabel style={font=\scriptsize},
  xtick={1,2,3,4,5},
  xticklabels={%
    \shortstack{AirWatch\\Norm.},
    \shortstack{AirWatch\\Alerting},
    \shortstack{AirWatch\\Cloud},
    \shortstack{REC\\Backbone},
    \shortstack{REC\\TEANS}},
  xticklabel style={font=\scriptsize, align=center},
  xmin=0.4, xmax=5.6,
  ymajorgrids=true,
]
  \addplot[aw box, boxplot prepared={
    draw position=1,
    lower whisker=1.45, lower quartile=3.06,
    median=4.30, upper quartile=5.43, upper whisker=8.07,
  }] coordinates {};

  \addplot[aw box, boxplot prepared={
    draw position=2,
    lower whisker=2.44, lower quartile=2.73,
    median=3.01, upper quartile=3.44, upper whisker=4.15,
  }] coordinates {};

  \addplot[aw box, boxplot prepared={
    draw position=3,
    lower whisker=1448, lower quartile=2393,
    median=2425, upper quartile=2458, upper whisker=2611,
  }] coordinates {};

  \addplot[rec box, boxplot prepared={
    draw position=4,
    lower whisker=16.11, lower quartile=17.03,
    median=17.95, upper quartile=19.11, upper whisker=19.87,
  }] coordinates {};

  \addplot[rec box, boxplot prepared={
    draw position=5,
    lower whisker=0.36, lower quartile=0.37,
    median=0.38, upper quartile=0.40, upper whisker=0.42,
  }] coordinates {};

\end{axis}

\node[anchor=north] at ($(panelA.south)!0.55!(panelB.south)+(0,-0.45cm)$) {%
  \begin{tikzpicture}[baseline]
    \fill[cyan!15!white] (0,0.04) rectangle (0.38,0.24); \draw[teal!70!black,thick] (0,0.04) rectangle (0.38,0.24);
    \node[font=\footnotesize,anchor=west] at (0.42,0.14) {AirWatch};
    \fill[orange!15!white] (2.6,0.04) rectangle (2.98,0.24); \draw[orange!70!black,thick] (2.6,0.04) rectangle (2.98,0.24);
    \node[font=\footnotesize,anchor=west] at (3.02,0.14) {REC};
    \draw[gray!60,dashed,thin] (4.9,0.14) -- (5.28,0.14);
    \node[font=\footnotesize,anchor=west] at (5.32,0.14) {Mean $\mu$};
    \draw[gray!60] (7.4,0.14) circle (0.09);
    \node[font=\footnotesize,anchor=west] at (7.52,0.14) {Outlier (IQR)};
  \end{tikzpicture}%
};

\end{tikzpicture}}
\caption{Latency distributions across the RASP deployment. Panel (a) shows the Edge$\to$Fog ingestion determinism for both REC and AirWatch ($\sim$65~ms). Panel (b) uses a log scale to compare all pipeline stages, from sub-millisecond TEANS decisions to multi-second Spark aggregation.}
    \label{fig:boxplots}
    \vspace{-0.3cm}
\end{figure*}
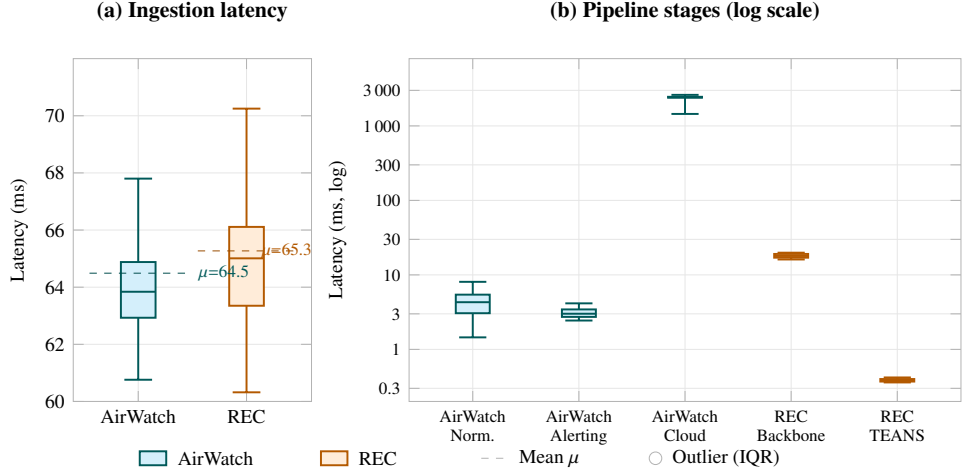

As summarized in the upper half of Table~\ref{tab:results}, latency remains fast and stable: the Edge$\to$Fog delay is noticeably higher in RASP than in POD (a $+52.6\%$ overhead), while the full Fog$\to$Cloud backbone path shows comparable latency across the two configurations. This consistency confirms that the multi-hop overhead is confined to the Edge$\to$Fog link and does not propagate to the inter-site backbone, reinforcing the attribution of latency variations to the RASP access path (\textit{Q3}). The Cloud decision logic (TEANS) was extremely responsive in both configurations, with processing times an order of magnitude below the ingestion stage, as detailed in the table. These results address \textit{Q2}, as the architecture successfully preserves provenance and timing evidence at every stage: each record carries a generation timestamp ($t_{gen}$), a Fog-arrival timestamp ($t_{arr}$), and a Cloud-ingestion marker, enabling post-hoc reconstruction of the full data path and stage-level latency attribution for any individual observation. The distribution across all tiers is illustrated in Fig.~\ref{fig:boxplots}, showing both ingestion stability (panel a) and the order-of-magnitude differences between stages (panel b).

\subsection{Use Case 2: AirWatch}
AirWatch stresses the continuum as a monitoring pipeline, where Fog services ingest environmental records via MQTT, normalize them, and forward them via Kafka toward cloud-side Spark aggregation.

Both configurations maintained a strict 5~s sampling cadence. As reported in the lower half of Table~\ref{tab:results}, the composite Edge$\to$Fog ingestion delay is consistently higher in RASP than in POD (a $+56.6\%$ overhead), closely mirroring the ingestion gap observed in the REC scenario. Fog-side normalization and alerting stages were highly efficient and essentially unaffected by the edge realization, with comparable latencies across configurations after removing transient network outliers ($<1.5\times$ IQR). These figures confirm, consistently with the REC results, that the RASP overhead is confined to the Edge$\to$Fog hop and does not propagate to the backbone or downstream Fog services.

Spark generated 14 aggregation windows per run: the 10 core windows of the experiment plus 4 additional windows required to flush the end-of-run backlog and watermark state. Cloud-side processing latency, measured directly from the Spark driver logs, was statistically comparable across configurations despite being the largest absolute figures in Table~\ref{tab:results}, reflecting the shared multi-tenant Cloud infrastructure noted in the table. Aggregated results are persisted to a dedicated cloud-tier InfluxDB instance, ensuring that historical monitoring data remains co-located with the analytics engine to optimize dashboarding and long-term storage tasks. These values answer \textit{Q4}, showing that the monitoring pipeline achieves stable throughput and low-jitter ingestion while maintaining deterministic aggregation behavior across independent repetitions.

\subsection{Cross-Scenario Interpretation}
The experimental results demonstrate that the same architectural substrate can host both control-oriented (REC) and monitoring-oriented (AirWatch) workloads with the same deployment primitives, thereby confirming \textit{Q1}. Table~\ref{tab:results} provides the compact quantitative comparison between virtualized and physical edge deployments, summarizing the performance profile that researchers can expect when deploying Cyber-Physical workflows over the SLICES continuum.

\begin{table}[t]
\centering
\caption{Aggregate performance comparison between virtualized (POD) and physical (RASP) edge tiers. 
         All latencies in milliseconds (ms). Values derived from 40 automated runs (10 per scenario and configuration).}
\label{tab:results}
\scriptsize
\setlength{\tabcolsep}{2.5pt}
\begin{threeparttable}
\begin{tabular*}{\columnwidth}{@{\extracolsep{\fill}}lccc@{}}
\toprule
\textbf{Pipeline Stage} &
  {\textbf{POD (Mean$\pm\sigma$)}} &
  {\textbf{RASP (Mean$\pm\sigma$)}} &
  {\textbf{Overhead}} \\
\midrule
\multicolumn{4}{@{}l}{\textit{AirWatch scenario (monitoring workload)}} \\
\addlinespace[2pt]
\quad Ingestion (Edge$\to$Fog)    & 41.19$\pm$0.93  & 64.49$\pm$4.06  & +56.6\% \\
\quad Normalization (Fog)         &  4.41$\pm$2.00  &  4.61$\pm$1.87  & -- \\
\quad Alerting (Fog sink)         &  3.30$\pm$0.55  &  3.11$\pm$0.47  & -- \\
\quad Cloud Aggregation (Spark)   & 2544$\pm$238   & 2347$\pm$203   & -- \\
\addlinespace[4pt]
\multicolumn{4}{@{}l}{\textit{REC scenario (control workload)}} \\
\addlinespace[2pt]
\quad Ingestion (Edge$\to$Fog)    & 42.77$\pm$2.72  & 65.27$\pm$4.47  & +52.6\% \\
\quad Backbone (Fog$\to$Cloud)    & 20.96$\pm$9.02  & 18.69$\pm$2.98  & -- \\
\quad Cloud Decision (TEANS)      &  0.391$\pm$0.015 &  0.384$\pm$0.021 & -- \\
\bottomrule
\end{tabular*}
\begin{tablenotes}[flushleft]
  \footnotesize
  \item[*] Overhead calculated as $(RASP/POD - 1)$. Reported values are filtered for transient outliers ($<1.5 \times$ IQR) to represent steady-state performance. Differences in Fog/Cloud stages are negligible as they reflect shared multi-tenant infrastructure.
\end{tablenotes}
\end{threeparttable}
\end{table}

The comparison shows that the architecture provides a unified message-driven foundation. While a physical edge overhead is visible in both workloads, its impact remains bounded and predictable, supporting the feasibility of real-time interactions over a distributed SLICES continuum.

\subsection{Threats to Validity and Limitations}
\label{sec:limitations}
The experimental results are subject to several limitations that define the scope of the current validation. First, the evaluation focuses on a specific three-site topology (Messina--Bologna-- Belgium (the selected SLICES site)); while representative of a distributed continuum, the results may vary under arbitrary network conditions or different inter-site latencies, for example, the $\pm$9.02~ms variance on the REC backbone path, while contained, reflects transient inter-site conditions that may differ across deployments. Second, while the campaign systematically compares physical and virtualized edges, it does not explore variations in component placement (e.g., moving anomaly detection from Fog to Cloud), which remains a key objective for future benchmarking studies. Third, the use of synthetic domain traces (e.g., CO$_2$ levels and energy profiles) allows for controlled experimentation but does not capture the full unpredictability of real-world sensor noise and field-data irregularities. Finally, the 10-minute experimental window provides a clear view of steady-state pipeline performance, but does not observe long-term phenomena such as resource fragmentation, software aging, or cumulative clock drift, for which the zero-variance record counts.
\section{Conclusion}

This paper presented a two-level reference architecture for Cloud Continuum experimentation built on the SLICES Cloud Continuum Blueprint. The first level separates the research-infrastructure substrate, which exposes and manages federated distributed resources, from the application-level Edge--Fog--Cloud pattern instantiated on top of it. The second level organizes Cyber--Physical workflows into stage-aware decompositions that make placement, timing, and provenance explicit. Two representative use cases, Renewable Energy Community management and AirWatch urban monitoring, were used to instantiate this architecture as a control-oriented and a monitoring-oriented workload, respectively, and both were validated through a systematic campaign of 40 automated runs comparing virtualized (POD) and physical edge (RASP) deployments.

The evaluation showed that the same architectural primitives and deployment descriptors can host workloads with fundamentally different timing, aggregation, and coordination requirements. In the REC scenario, the architecture maintained sub-90~ms end-to-end latency even on physical hardware, preserved provenance across the full Edge--Fog--Cloud path, and produced consistent Digital Twin decisions across all configurations. In the AirWatch scenario, it sustained deterministic ingestion, stable fog-side normalization and alerting, and reproducible cloud-side aggregation across independent repetitions. The physical edge overhead was measurable to the Edge--Fog hop, and did not propagate to the backbone or to cloud-side processing, indicating that the architecture can bound the impact of heterogeneous edge resources in a predictable way.
The broader implication is that cloud-continuum experimentation benefits from being organized around reusable workflow evidence rather than around single-deployment benchmarks. The same physical and virtual resources can yield different performance profiles depending on workload semantics, and the infrastructure must therefore expose enough per-stage information to interpret those differences. By combining deployment repeatability, stage-level latency decomposition, message provenance, and deterministic aggregation, the proposed approach provides a shared experimental basis for comparing future continuum applications under controlled and reproducible assumptions.

{
Future work will extend the evaluation along three directions.
The first is the experimental validation of the HPC back-end stage already specified in the workflow decomposition. The REC optimization and model-calibration stage, together with the AirWatch long-term sensitivity analysis, will be offloaded to batch-oriented back-end resources, measuring the batch duration, data-transfer overhead, and model-output availability defined in Table~\ref{tab:workflow-decomposition}. This will allow the trade-off between offloading gains and continuum responsiveness to be quantified using the same descriptor-based method adopted in this paper.
The second direction concerns scale and coverage: more sites, higher sampling rates, longer observation windows, and additional workload profiles stressing placement sensitivity and recovery behavior. The third is the consolidation of the artifact package through run descriptors, configuration snapshots, and raw log archives, so that the experimental scenarios can be replayed and evolved by other researchers. These extensions will move the proposed approach from a validated architectural proposal toward a reusable benchmark methodology for cloud-continuum workflows, one that evaluates not only throughput and average latency, but also provenance completeness, placement sensitivity, and the distinction between physical and virtualized edge resources.}

\section*{Acknowledgements}
This work was partial funding from the European Union’s Horizon Europe Research and Innovation Programme under Grant Agreement No. 101287692 (SLICES-IP).

\section*{CRediT authorship contribution statement}
\noindent\textbf{Fabio Orazio Mirto:} Investigation, Software, Validation, Data Curation, Visualization, Writing - Original Draft.\\
\textbf{Giuseppe Tricomi:} Investigation, Methodology, Validation, Writing - Original Draft, Visualization.\\
\textbf{Luca D'Agati:} Investigation, Methodology, Writing - Original Draft, Visualization.\\
\textbf{Andrea Sabbioni:} Investigation, Methodology, Software, Supervision.\\
\textbf{Stefano Silvestri:} Supervision, Funding acquisition.\\
\textbf{Francesco Longo:} Investigation, Writing – Review \& Editing, Supervision.\\
\textbf{Giovanni Merlino:} Conceptualization, Writing – Review \& Editing, Supervision.\\
\textbf{Armir Bujari:} Methodology, Writing – review \& editing, Supervision.\\
\textbf{Paolo Bellavista:} Funding acquisition, Project administration.\\
\textbf{Antonio Puliafito:} Funding acquisition, Project administration.\\

\section*{Declaration of competing interest}
The authors declare that they have no known competing financial interests or personal relationships that could have appeared to influence the work reported in this paper.

\section*{Declaration of generative AI and AI-assisted technologies in the manuscript preparation process}
During the preparation of this work, the authors used GenAI to improve the clarity and accuracy of the language. Having used this service, they reviewed and edited the content as necessary and take full responsibility for the published article.

{\small
\bibliographystyle{elsarticle-num}
\bibliography{references}

@article{planetlab,
  author  = {Chun, Brent and Culler, David and Roscoe, Timothy and Bavier, Andy and Peterson, Larry and Wawrzoniak, Mike and Bowman, Mic},
  title   = {{PlanetLab}: An overlay testbed for broad-coverage services},
  journal = {Computer Communication Review},
  volume  = {33},
  number  = {3},
  pages   = {3--12},
  year    = {2003},
  doi     = {10.1145/956993.956995}
}

@article{geni,
  author  = {Berman, Mark and Chase, Jeffrey S. and Landweber, Larry and Nakao, Akihiro and Ott, Max and Raychaudhuri, Dipankar and Ricci, Robert and Seskar, Ibrahim},
  title   = {{GENI}: A federated testbed for innovative network experiments},
  journal = {Computer Networks},
  volume  = {61},
  pages   = {5--23},
  year    = {2014},
  doi     = {10.1016/j.bjp.2013.12.037}
}

@INPROCEEDINGS{iotlab,
  author={Adjih, Cedric and Baccelli, Emmanuel and Fleury, Eric and Harter, Gaetan and Mitton, Nathalie and Noel, Thomas and Pissard-Gibollet, Roger and Saint-Marcel, Frederic and Schreiner, Guillaume and Vandaele, Julien and Watteyne, Thomas},
  booktitle={2015 IEEE 2nd World Forum on Internet of Things (WF-IoT)}, 
  title={FIT IoT-LAB: A large scale open experimental IoT testbed}, 
  year={2015},
  volume={},
  number={},
  pages={459-464},
  doi={10.1109/WF-IoT.2015.7389098}}

@inproceedings {chameleon,
author = {Kate Keahey and Jason Anderson and Zhuo Zhen and Pierre Riteau and Paul Ruth and Dan Stanzione and Mert Cevik and Jacob Colleran and Haryadi S. Gunawi and Cody Hammock and Joe Mambretti and Alexander Barnes and Fran{\c c}ois Halbah and Alex Rocha and Joe Stubbs},
title = {Lessons Learned from the Chameleon Testbed},
booktitle = {2020 USENIX Annual Technical Conference (USENIX ATC 20)},
year = {2020},
isbn = {978-1-939133-14-4},
pages = {219--233},
url = {https://www.usenix.org/conference/atc20/presentation/keahey},
publisher = {USENIX Association},
month = jul
}

@inproceedings{edgenet,
  author    = {Senel, Berk Can and Mouchet, Maxime and Cappos, Justin and Fourmaux, Olivier and Friedman, Timur and McGeer, Rick},
  title     = {{EdgeNet}: A Multi-Tenant and Multi-Provider Edge Cloud},
  booktitle = {Proceedings of the 4th International Workshop on Edge Systems, Analytics and Networking},
  pages     = {49--54},
  year      = {2021},
  doi       = {10.1145/3434770.3459737}
}

@misc{slicesds,
  author       = {{European Commission}},
  title        = {{SLICES-DS}: Scientific Large-scale Infrastructure for Computing/Communication Experimental Studies -- Design Study},
  year         = {2020},
  note         = {CORDIS project fact sheet, Grant Agreement No. 951850},
  doi          = {10.3030/951850}
}

@misc{crossplane,
  author       = {{Crossplane Project}},
  title        = {Crossplane Documentation},
  year         = {2024},
  howpublished = {\url{https://docs.crossplane.io}},
  note         = {Cloud Native Computing Foundation (CNCF) project documentation},
  urldate      = {2024}
}

@INPROCEEDINGS{stack4things_longo2014,
  author={Longo, Francesco and Bruneo, Dario and Distefano, Salvatore and Merlino, Giovanni and Puliafito, Antonio},
  booktitle={2015 3rd International Conference on Future Internet of Things and Cloud}, 
  title={Stack4Things: An OpenStack-Based Framework for IoT}, 
  year={2015},
  volume={},
  number={},
  pages={204-211},
  doi={10.1109/FiCloud.2015.97}}

@inproceedings{fogrole,
  author    = {Bonomi, Flavio and Milito, Rodolfo and Zhu, Jiang and Addepalli, Sateesh},
  title     = {Fog Computing and Its Role in the Internet of Things},
  booktitle = {Proceedings of the First Edition of the {MCC} Workshop on Mobile Cloud Computing},
  pages     = {13--16},
  year      = {2012},
  doi       = {10.1145/2342509.2342513}
}

@article{cloudtofog,
  author  = {Osanaiye, Olakunle and Chen, Shuhao and Yan, Zheng and Lu, Rongxing and Choo, Kim-Kwang Raymond and Dlodlo, Mlamuli},
  title   = {From Cloud to Fog Computing: A Review and a Conceptual Live {VM} Migration Framework},
  journal = {IEEE Access},
  volume  = {5},
  pages   = {8284--8300},
  year    = {2017},
  doi     = {10.1109/ACCESS.2017.2692960}
}

@article{fogiot,
  author  = {Chiang, Mung and Zhang, Tao},
  title   = {Fog and {IoT}: An Overview of Research Opportunities},
  journal = {IEEE Internet of Things Journal},
  volume  = {3},
  number  = {6},
  pages   = {854--864},
  year    = {2016},
  doi     = {10.1109/JIOT.2016.2584538}
}

@article{edgevision,
  author  = {Shi, Weisong and Cao, Jie and Zhang, Quan and Li, Youhuizi and Xu, Lanyu},
  title   = {Edge Computing: Vision and Challenges},
  journal = {IEEE Internet of Things Journal},
  volume  = {3},
  number  = {5},
  pages   = {637--646},
  year    = {2016},
  doi     = {10.1109/JIOT.2016.2579198}
}

@article{edgeemergence,
  author  = {Satyanarayanan, Mahadev},
  title   = {The Emergence of Edge Computing},
  journal = {Computer},
  volume  = {50},
  number  = {1},
  pages   = {30--39},
  year    = {2017},
  doi     = {10.1109/MC.2017.9}
}

@article{iotfogcloud, 
    title = {The Internet of Things, Fog and Cloud continuum: Integration and challenges}, 
    journal = {Internet of Things}, 
    volume = {3-4}, 
    pages = {134-155}, 
    year = {2018}, 
    issn = {2542-6605}, 
    doi = {10.1016/j.iot.2018.09.005},
    author = {Luiz Bittencourt and Roger Immich and Rizos Sakellariou and Nelson Fonseca and Edmundo Madeira and Marilia Curado and Leandro Villas and Luiz DaSilva and Craig Lee and Omer Rana}
}

@article{fogresource,
  author  = {Hong, Cheol-Ho and Varghese, Blesson},
  title   = {Resource Management in Fog/Edge Computing: A Survey on Architectures, Infrastructure, and Algorithms},
  journal = {ACM Computing Surveys},
  volume  = {52},
  number  = {5},
  pages   = {1--37},
  year    = {2019},
  doi     = {10.1145/3326066}
}

@Inbook{orchestrationedge,
editor="Lynn, Theo and Mooney, John G. and Lee, Brian and Endo, Patricia Takako",
title="Orchestration from the Cloud to the Edge",
booktitle="The Cloud-to-Thing Continuum: Opportunities and Challenges in Cloud, Fog and Edge Computing",
year="2020",
publisher="Springer International Publishing",
address="Cham",
chapter="4",
pages="61--77",
isbn="978-3-030-41110-7",
doi="10.1007/978-3-030-41110-7_4"
}

@article{dtmanufacturing,
  author  = {Kritzinger, Werner and Karner, Matthias and Traar, Georg and Henjes, Jan and Sihn, Wilfried},
  title   = {Digital Twin in manufacturing: A categorical literature review and classification},
  journal = {IFAC-PapersOnLine},
  volume  = {51},
  number  = {11},
  pages   = {1016--1022},
  year    = {2018},
  doi     = {10.1016/j.ifacol.2018.08.474}
}

@article{dtcharacterising,
  author  = {Jones, David and Snider, Chris and Nassehi, Aydin and Yon, Jilani and Hicks, Ben},
  title   = {Characterising the Digital Twin: A systematic literature review},
  journal = {CIRP Journal of Manufacturing Science and Technology},
  volume  = {29},
  pages   = {36--52},
  year    = {2020},
  doi     = {10.1016/j.cirpj.2020.02.002}
}

@article{dtenabling,
  author  = {Fuller, Aidan and Fan, Zhong and Day, Charles and Barlow, Chris},
  title   = {Digital Twin: Enabling Technologies, Challenges and Open Research},
  journal = {IEEE Access},
  volume  = {8},
  pages   = {108952--108971},
  year    = {2020},
  doi     = {10.1109/ACCESS.2020.2998358}
}

@article{smartmeterarch,
  author  = {Oprea, Simona-Vasilica and B{\^a}ra, Adela},
  title   = {An Edge-Fog-Cloud computing architecture for {IoT} and smart metering data},
  journal = {Peer-to-Peer Networking and Applications},
  volume  = {16},
  pages   = {1415--1430},
  year    = {2023},
  doi     = {10.1007/s12083-022-01436-y}
}

@article{energycommunities,
  author={Oprea, Simona-Vasilica and B{\^a}ra, Adela},
  title= {Edge and fog computing using {IoT} for direct load optimization and control with flexibility services for citizen energy communities},
  journal= {Knowledge-Based Systems},
  volume= {228},
  pages= {107293},
  year= {2021},
  doi= {10.1016/j.knosys.2021.107293}
}

@article{recpackage,
  author  = {Lowitzsch, Jens and Hoicka, Christina E. and van Tulder, Frank J.},
  title   = {Renewable Energy Communities under the 2019 European Clean Energy Package --- Governance Model for the Energy Clusters of the Future?},
  journal = {Renewable and Sustainable Energy Reviews},
  volume  = {122},
  pages   = {109489},
  year    = {2020},
  doi     = {10.1016/j.rser.2019.109489}
}

@article{p2pmarkets,
title = {Peer-to-peer and community-based markets: A comprehensive review},
journal = {Renewable and Sustainable Energy Reviews},
volume = {104},
pages = {367-378},
year = {2019},
issn = {1364-0321},
doi = {10.1016/j.rser.2019.01.036},
author = {Tiago Sousa and Tiago Soares and Pierre Pinson and Fabio Moret and Thomas Baroche and Etienne Sorin},
}

@article{airqualitysensing,
title = {Applications of low-cost sensing technologies for air quality monitoring and exposure assessment: How far have they gone?},
journal = {Environment International},
volume = {116},
pages = {286-299},
year = {2018},
issn = {0160-4120},
doi = {10.1016/j.envint.2018.04.018},

author = {Lidia Morawska and Phong K. Thai and Xiaoting Liu and Akwasi Asumadu-Sakyi and Godwin Ayoko and Alena Bartonova and Andrea Bedini and Fahe Chai and Bryce Christensen and Matthew Dunbabin and Jian Gao and Gayle S.W. Hagler and Rohan Jayaratne and Prashant Kumar and Alexis K.H. Lau and Peter K.K. Louie and Mandana Mazaheri and Zhi Ning and Nunzio Motta and Ben Mullins and Md Mahmudur Rahman and Zoran Ristovski and Mahnaz Shafiei and Dian Tjondronegoro and Dane Westerdahl and Ron Williams},
}

@INPROCEEDINGS {dagati2025iotorchestration,
author = { D'Agati, Luca and Tricomi, Giuseppe and Arena, Michele and Longo, Francesco and Puliafito, Antonio and Merlino, Giovanni },
booktitle = { 2025 IEEE 25th International Symposium on Cluster, Cloud and Internet Computing Workshops (CCGridW) },
title = {{ IoT Orchestration in the Compute Continuum: Integrating Kubernetes with Stack4Things }},
year = {2025},
volume = {},
ISSN = {},
pages = {140-147},
url = {https://doi.ieeecomputersociety.org/10.1109/CCGridW65158.2025.00028}
}

@article{cicceri2023rec,
  author  = {Cicceri, Giovanni and Tricomi, Giuseppe and D'Agati, Luca and
             Longo, Francesco and Merlino, Giovanni and Puliafito, Antonio},
  title   = {A Deep Learning-Driven Self-Conscious Distributed
             Cyber-Physical System for Renewable Energy Communities},
  journal = {Sensors},
  volume  = {23},
  number  = {9},
  pages   = {4549},
  year    = {2023},
  doi     = {10.3390/s23094549}
}

@inproceedings{tricomi2024urban,
  author    = {Tricomi, Giuseppe and D'Agati, Luca and Longo, Francesco and
               Merlino, Giovanni and Puliafito, Antonio and Silvestri, Stefano},
  title     = {Paving the Way for an Urban Intelligence
               {OpenStack}-Based Architecture},
  booktitle = {2024 IEEE International Conference on Smart Computing
               (SMARTCOMP)},
  pages     = {284--289},
  year      = {2024},
  doi       = {10.1109/SMARTCOMP61445.2024.00069}
}

@article{faasiot,
title = {FaaS for IoT: Evolving Serverless towards Deviceless in I/Oclouds},
journal = {Future Generation Computer Systems},
volume = {154},
pages = {189-205},
year = {2024},
issn = {0167-739X},
doi = {10.1016/j.future.2023.12.029},
author = {Giovanni Merlino and Giuseppe Tricomi and Luca D’Agati and Zakaria Benomar and Francesco Longo and Antonio Puliafito}
}

@article{airqualitycommercial,
title = {Can commercial low-cost sensor platforms contribute to air quality monitoring and exposure estimates?},
journal = {Environment International},
volume = {99},
pages = {293-302},
year = {2017},
issn = {0160-4120},
doi = {10.1016/j.envint.2016.12.007},
author = {Nuria Castell and Franck R. Dauge and Philipp Schneider and Matthias Vogt and Uri Lerner and Barak Fishbain and David Broday and Alena Bartonova},
}

@article{airqualitycalibration,
  author  = {Maag, Blaise and Zhou, Zimu and Thiele, Lothar},
  title   = {A Survey on Sensor Calibration in Air Pollution Monitoring Deployments},
  journal = {IEEE Internet of Things Journal},
  volume  = {5},
  number  = {6},
  pages   = {4857--4870},
  year    = {2018},
  doi     = {10.1109/JIOT.2018.2853660}
}

@article{airqualityfog,
  author  = {Bharathi, B. and Rafeeq, A. H. M. and Prakash, M.},
  title   = {Fog computing enabled air quality monitoring and prediction leveraging deep learning in {IoT}},
  journal = {Journal of Intelligent \& Fuzzy Systems},
  volume  = {43},
  number  = {1},
  pages   = {1501--1512},
  year    = {2022},
  doi     = {10.3233/JIFS-212713}
}

@article{workflowescience,
  author  = {Deelman, Ewa and Gannon, Dennis and Shields, Matthew and Taylor, Ian},
  title   = {Workflows and e-Science: An Overview of Workflow System Features and Capabilities},
  journal = {Future Generation Computer Systems},
  volume  = {25},
  number  = {5},
  pages   = {528--540},
  year    = {2009},
  doi     = {10.1016/j.future.2008.06.012}
}

@inproceedings{provenanceworkflows,
  author    = {Davidson, Susan B. and Freire, Juliana},
  title     = {Provenance and Scientific Workflows: Challenges and Opportunities},
  booktitle = {Proceedings of the 2008 {ACM} {SIGMOD} International Conference on Management of Data},
  pages     = {1345--1350},
  year      = {2008},
  doi       = {10.1145/1376616.1376772}
}

@article{computingcontinuumworkflows,
  author  = {Balouek-Thomert, Daniel and Renart, Eduard Gibert and Zamani, Ali Reza and Simonet, Anthony and Parashar, Manish},
  title   = {Towards a Computing Continuum: Enabling Edge-to-Cloud Integration for Data-Driven Workflows},
  journal = {The International Journal of High Performance Computing Applications},
  volume  = {33},
  number  = {6},
  pages   = {1159--1174},
  year    = {2019},
  doi     = {10.1177/1094342019877383}
}

@inproceedings{e2clab,
  author    = {Rosendo, Daniel and Silva, Pedro and Simonin, Matthieu and Costan, Alexandru and Antoniu, Gabriel},
  title     = {{E2Clab}: Exploring the Computing Continuum through Repeatable, Replicable and Reproducible Edge-to-Cloud Experiments},
  booktitle = {Proceedings of the 2020 {IEEE} International Conference on Cluster Computing ({CLUSTER})},
  pages     = {176--186},
  year      = {2020},
  doi       = {10.1109/CLUSTER49012.2020.00028}
}

@article{streamflow,
  author  = {Colonnelli, Iacopo and Cantalupo, Barbara and
             Merelli, Ivan and Aldinucci, Marco},
  title   = {{StreamFlow}: cross-breeding Cloud with {HPC}},
  journal = {IEEE Transactions on Emerging Topics in Computing},
  volume  = {9},
  number  = {4},
  pages   = {1723--1737},
  year    = {2021},
  doi     = {10.1109/TETC.2020.3019202}
}

@article{pegasus,
  author  = {Deelman, Ewa and Vahi, Karan and Juve, Gideon and
             Rynge, Mats and Callaghan, Scott and
             Maechling, Philip J. and Mayani, Rajiv and
             Chen, Weiwei and Ferreira da Silva, Rafael and
             Livny, Miron and Wenger, Kent},
  title   = {Pegasus, a workflow management system for science
             automation},
  journal = {Future Generation Computer Systems},
  volume  = {46},
  pages   = {17--35},
  year    = {2015},
  doi     = {10.1016/j.future.2014.10.008}
}
}
\end{document}